# Large-angle OBServaTory with Energy Resolution for Synoptic X-ray Studies (LOBSTER-SXS)

Paul Gorenstein[*],
Harvard-Smithsonian Center for Astrophysics, 60 Garden St. MS-4, Cambridge, MA USA 02138

## ABSTRACT

The soft X-ray band hosts a larger, more diverse range of variable sources than any other region of the electromagnetic spectrum. They are stars, compact binaries, SMBH's, the X-ray components of Gamma-Ray Bursts, their X-ray afterglows, and soft X-ray flares from supernova. We describe a concept for a very wide field (~ 4 ster) modular hybrid X-ray telescope system that can measure positions of bursts and fast transients with as good as arc second accuracy, the precision required to identify fainter and increasingly more distant events. The dimensions and materials of all telescope modules are identical. All but two are part of a cylindrical lobster-eye telescope with flat double sided mirrors that focus in one dimension and utilize a coded mask for resolution in the other. Their positioning accuracy is about an arc minute. The two remaining modules are made from the same materials but configured as a Kirkpatrick-Baez telescope with longer focal length that focuses in two dimensions. When pointed it refines the hybrid telescope's arc minute positions to an arc second and provides larger effective area for spectral and temporal measurements. Above 10 keV the mirrors act as an imaging collimator with positioning capability. For short duration events this hybrid focusing/coded mask system is more sensitive and versatile than either a 2D coded mask or a 2D lobster-eye telescope. Very wide field X-ray telescopes have become feasible as the ability to fabricate large area arrays of CCD and CMOS detectors has improved. This instrument's function in the soft X-ray band is similar to that of Swift in hard X-ray band and there are a greater variety of fast transients in the soft X-ray band. An instrument with considerably more sensitivity than current wide field X-ray detectors would be compatible with a modest NASA Explorer mission.

**Keywords:** X-ray telescopes, lobster-eye telescopes, X-ray variability

## 1. INTRODUCTION

### 1.1. X-ray Variability

In all bands of the electromagnetic spectrum there is interest in developing a new generation of synoptic astronomical telescopes. Pan-STARRS and the Large Synoptic Survey Telescope (LSST) will serve the optical band[1]. The Square Kilometer Array will be the principal radio facility. The coded mask detectors of **Swift** (**BAT**)[2] cover the hard X-ray band. The **Fermi Large Area Telescope (LAT)**[3] is a synoptic gamma-ray facility. However, temporal variability is most remarkable in the soft X-ray band. Within our galaxy the time scales range from millisecond pulsars to transient source outbursts at decadal intervals. Observing changes in their intensities and spectra are among the few means of studying black holes, neutron stars and testing general relativity. Randomly occurring X-ray flares emanate from young and pre-main sequence stellar clusters. The extragalactic sky contains the most rapidly varying sources, gamma-ray bursts (GRBs) and X-ray flashes that exist for intervals ranging from milliseconds to minutes. A 400 second long soft X-ray flare from a supernova explosion that was detected serendipitously in the field of the **Swift** **XRT** is the most recent discovery[4]. At the other extreme, changes in the intensity of an AGN can occur over a time span of months to years, reflecting changes in jet activity or possibly triggered by the central SMBH tidally disrupting and accreting a star[5], or being obscured by an orbiting cloud[6].

The importance of observing temporal variations in the X-ray band and advocating developing the next generation of instrumentation to do so are subjects of White Papers submitted to the NAS 2010 Astronomy Decadel Survey Committee. In their paper entitled "Scientific Productivity with X-ray All-Sky Monitors"[7] Remillard, Levine, and McClintock evaluate the results of a decade and a half of operations of the **Rossi X-ray Timing Explorer** and offer compelling reasons for developing a more advanced X-ray all-sky monitor than the **ASM** of **RXTE**, and Japan's **MAXI** (installed on the Japanese Module of the International Space Station). Addressing extra-galactic as well as galactic temporal variations, a large community of authors submitted a White Paper, entitled "The Dynamic X-ray Sky of the Local Universe"[8]. Their key goals include detecting supernova at the time of explosion, observing stars with active coronas, detecting quiescent supermassive black holes and "exploring new phase space of the Transient Universe at X-ray wavelengths". They refer specifically to the ***EXIST*** mission concept[9] as satisfying those objectives in the hard X-ray band. The discovery of soft X-ray flares from supernova, the existence of "X-ray flashes", X-ray afterglows from GRBs, and possibly, "orphan afterglows" show that those goals apply at least equally well to the soft X-ray band.

---

[*] goren@cfa.harvard.edu; phone 1 617 495-7250; fax 1 617 495-7356

A soft X-ray detector with good angular resolution, high sensitivity and a very large field of view is required to explore this important domain. Scanning X-ray detectors with a limited field of view such as the **RXTE** ASM and **MAXI** miss many short lived transients and obtain only intermittent samples of those that are longer lived. **LAT**, the wide field coded mask detector aboard *Swift*, is the only wide field telescope option for the hard X-ray/soft gamma-ray bands. In the soft X-ray band the diffuse cosmic X-ray background and the known sources are much more intense. Consequently the background level in a wide field coded mask is very high. Many of the known soft X-ray sources vary making it even more difficult for a coded mask to detect short lived transients. However, soft X-rays can be focused much more effectively than hard X-rays. Therefore a very wide field focusing telescope, which focuses all sources within its broad field of view onto different regions of the detector, is a more appropriate tool for synoptic studies of soft X-ray sources.

We describe a concept for a very wide field focusing/coded mask hybrid soft X-ray telescope with a large dynamic range that is capable of resolving individual transients in a field containing multiple variable objects. The nominal field of view is four steradians. It is highly modular; the field of view can be tailored to fit spacecraft mass and volume accommodations. The prime bandwidth for detecting transients and measuring positions is 0.5 to 6 keV. Above 6 keV where its X-ray reflectivity has diminished the lobster-eye telescope functions with less angular resolution and less sensitivity as an imaging collimator up to 15 keV or higher depending on the detector. The coded mask's functionality is not energy dependent.

**1.2 Telescope Description**

The telescope is composed of identical size modules made of the same materials. All but two are a hybrid of a lobster-eye telescope that images by focusing in one dimension and utilizes a coded mask for angular resolution in the other. It has less effective area than a two dimensional coded mask system of similar size, but is more sensitive because the background is much lower. Positions of transient events with arc minute or better precision would be available very rapidly, essentially instantaneously in the focusing dimension and very quickly in the other by limiting the region that need be correlated with the coded mask to only a small fraction of the sky, along only one dimension, and most likely only a single source within the region. Two remaining modules are configured as a Kirkpatrick-Baez (KB) telescope[10] that has larger effective area and finer resolution. Like *Swift*'s use of the **XRT** the spacecraft can point the KB modules immediately at that position to refine it to an arc second and improve the measurement of its spectrum. While the KB module is pointed at the object it can remain within the field of view of the lobster-eye.

A wide field telescope system with considerable sensitivity is compatible with a NASA Modest Explorer mission as described in the November, 2010 NASA Announcement of Opportunity.

This type optics was described by Schmidt in 1975[11] and the hybrid focusing/coded mask concept by Gorenstein and Mauche in 1984[12]. The importance of a very broad field of view synoptic X-ray telescope became much more evident following the discovery of X-ray afterglows of gamma-ray bursts, X-ray flashes, and most recently, the discovery of short lived X-ray flares from certain type supernova. At the same time there have been favorable developments on technical aspects such as the availability of low mass, low cost, high quality glass flats for the optics and progress in the development of detectors. Now there exists the capability to fabricate a large number of solid state position sensitive detectors with good energy resolution, such as CCDs, CMOS detectors, and silicon drift position sensitive solid state X-ray detectors[13,14,15]. A cylindrical, large area gas electron multiplier[16], which is an evolved position sensitive gas proportional counter is a lower cost alternative to an array of solid state detectors but not with as good spatial and energy resolution.

The expected performance of a modest instrument as determined by ray tracing is described in this paper.

# 2. FAST X-RAY TRANSIENT SOURCES

## 2.1. Gamma-ray Bursts, X-ray Flashes and X-ray Afterglows

The X-ray afterglows of GRBs are the key to finding their positions, optically identifying their host galaxies and measuring their redshifts[17]. In addition, there is a class of GRB-like events whose content is predominantly X-rays. *BeppoSAX*[18] and *HETE-2*[19] detected bursts they called "X-ray Flashes" (XRF) that resemble GRBs in many respects but whose spectra were much softer and resided mostly in the 2 to 30 keV X-ray band. In addition *HETE-2* observed an intermediate class of events that they named "X-Ray Rich" where the 2-30 keV fluence is comparable to the fluence above 30 keV. To avoid an inexplicable high luminosity it is generally agreed that GRBs are strongly beamed. The X-ray afterglow is most likely the product of a secondary shock wave or scattering and may not be as strongly beamed. Woods and Loeb[20], suggested that X-ray flashes are GRBs but viewed at a larger angle off the center of the beam. More sensitive all-sky exposures in the X-ray band may detect "orphan" X-ray afterglows[21]. They can be detected, identified and yield as much information about the event and host galaxy as a GRB.

GRBs with very high redshift are the subject of much attention because of their connection to star formation in the early universe. One of the most distant objects ever detected was the host galaxy of GRB 090423 with a redshift of 8.2[22, 23]. At large values of z the redshift increases very rapidly as a function of distance or look-back time. Therefore, we can expect that for even slightly longer distance than at z = 8.2 the intrinsic GRB spectrum would be softened considerably. The sensitivity of searches for ever more distant gamma-ray bursts is likely to depend increasingly upon detecting their X-ray component. GRBs will occur within the field of view of this instrument. The complete history of the X-ray afterglow will be observed starting from its birth

## 2.2. Supernova

Fortuitously the onset of a supernova explosion was detected by the *Swift* XRT during an exposure scheduled for another object in the galaxy NGC 2770[3]. A soft X-ray flare lasting about 400 seconds occurred at a position quite distant from the intended target. Its peak flux was $6.9 \times 10^{-10}$ ergs/cm$^{-2}$ s$^{-1}$ (0.3 to 10 keV), which is about 2 % of the Crab Nebula and relatively intense for an extragalaxtic X-ray source at a distance of 25 mpc. Its spectrum was a power law with a photon index of 2.3 ± 0.3, and a hydrogen column density of $6.9 \times 10^{21}$ cm$^{-2}$, in excess of the absorption within the Milky Way. Although the SN was within its field of view, **BAT,** the *Swift* hard X-ray detector, did not detect it.

Chevalier and Fransson[24] attribute the serendipitous discovery to the shock breakout emission from a normal Type Ib/c supernova. Soft X-ray flares are expected to occur also with some Type 2 supernova[25]. This view suggests that many soft X-ray flare events from supernova occur each year whose spectra are below the 15–150 keV band of the *Swift* BAT.

# 3. THE WIDE FIELD X-RAY TELESCOPE

## 3.1 Optics and coded mask

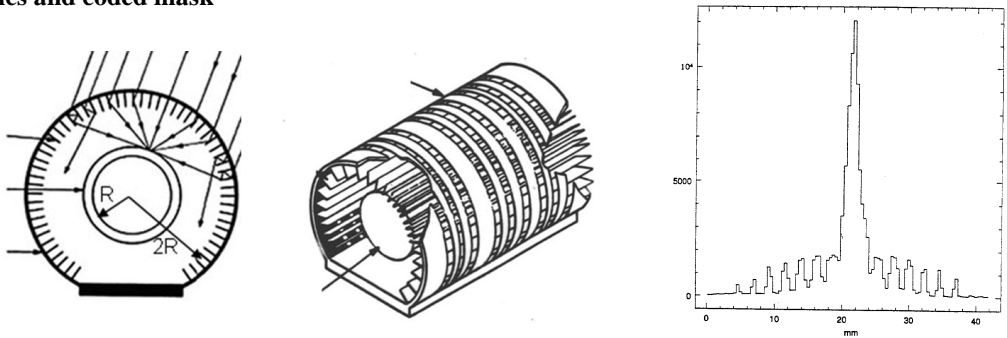

Figure 1. The left panel shows X-rays from a fixed direction that are focused on to a cylindrical position sensitive detector whose radius is half that of the reflectors. Some rays miss the reflectors and arrive at the detector along their original direction but their angular range is limited by the collimation action of the reflectors. The center

panel shows the coded mask surrounding the reflectors. The mask contains a pseudo random distribution of open slits along the axis of the cylinder. The right panel is an image obtained in the laboratory with a 1-D lobster-eye telescope.

The telescope shown in Fig. 1 focuses in one dimension to a line image. An exterior coded mask provides angular resolution in the other. Flat mirrors are evenly spaced around the circumference of a cylinder with their faces along radii. Both sides of the reflectors are coated with Ir or Pt plus a thin C overcoat. They act either as a mirror or as an imaging collimator depending upon whether the angle of incidence is above or below the critical angle. A spacecraft with a diameter of 2 m can accommodate a reflector array whose radius is 1 m, covers 180 degrees of azimuth, and is 1 m high. The cylindrical position sensitive detector array would have a radius of 50 cm and be 50 cm high. The lobster-eye/coded mask's field of view is about 4 ster (area > 0.35 maximum) plus more field of view with less effective area.

The telescope would be divided around the azimuth and along the axis of the cylinder into modules of the same size. Forty-eight, 25 cm x 25 cm modules, 12 along the azimuth and 4 along the axis would cover a 4 ster field. One position, is occupied by the front section of a Kirkpatrick-Baez (KB) telescope. The orthogonal rear section is immediately behind it. The KB modules are made from the same structural and reflector materials as the lobster-eye modules. The only differences between them are that the KB mirrors are coated on only one side, are not uniformly spaced, and are slightly curved into parabolas. The KB telescope images in two dimensions with a one degree field of view and has finer angular resolution and larger effective area below 5 keV than the lobster-eye modules. The studies on figure formation that have been carried out for NuSTAR and the large telescope of the International X-ray Observatory show that a telescope with a resolution much better than an arc minute can be constructed from thermally slumped glass[26]. When pointed at a transient event detected by the lobster-eye modules the KB modules can refine its position to about an arc second and measure its spectrum in more detail, including the low energy cutoff within the host galaxy of a GRB and provide more timing information.

Prototypes of lobster-eye telescopes consisting of glass flats were constructed and tested by Gorenstein et al[27] and by Hudec et al[28]. The former was made in one dimension. The latter was a two dimensional imager consisting of two orthogonal one dimensional arrays in series. Smooth, flat glass stock is commercially available with a large range of thickness.

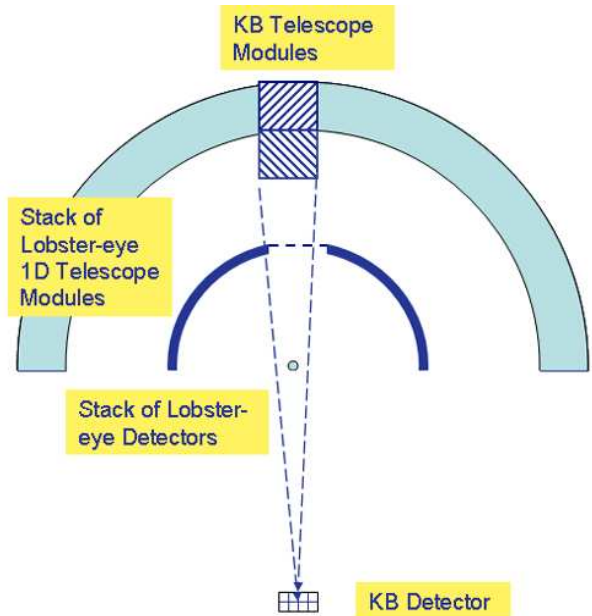

Figure 2. A position within the stack of hybrid telescope modules is occupied by two orthogonal focusing modules in series that form a Kirkpatrick-Baez telescope with a focal length of 1.9 m or the maximum allowed by the spacecraft envelope. The orientation of the KB telescope need not be as shown.

The dimensions and the substrate materials of the two modules that form a KB telescope will be similar to those of the hybrid telescope modules. Because their materials and assembly processes are similar the KB modules and the lobster-eye modules are not distinctly different telescopes as are for example, the **XRT** and **LAT** of *Swift*.

The curvature of the KB mirrors could be imparted by thermal slumping, a process used for NuSTAR[29] and laboratory studies for IXO. Figure formation of each KB mirror will be aided by constraints along the two supported edges.

The coded mask that provides angular resolution in the polar direction consists of open semi-circular slits up to 180 degrees along the azimuth of the partial cylinder and pseudo randomly spaced along the cylinder axis. The material would be 100 microns of titanium or nickel with 200 micron wide slits. It was not included in the ray tracing simulations but its transmission including its support structure of normally incident X-rays was assumed to be is 45% in estimating the effective area.

### 3.2 Determining positions

With the pointing direction fixed the process of finding a position for a burst or flare is initiated by detecting an increase in count rate by searching on several time scales simultaneously within 8 arc min regions of azimuth, which are the sum of two successive 4 arc min regions, i.e. (0-4 + 4-8), (4-8 + 8-12), (8-12 + 12-16), etc. The position of the centroid of the increase above the quiescent background profile will be found quickly. Correlating the coded mask with the event's axial coordinates within a several arc min region of azimuth containing the peak will quickly yield an arc min position in the polar direction of the transient plus in some cases the position of a known source that may be within the same region of azimuth. The arc minute or coarse position will be transmitted world-wide through the Gamma-ray Burst Coordination Network, International Supernovae Network, and others sites monitored by robotic and human observers. At the same time the KB modules will point to the transient and determine a fine position that is the order of an arc second and very quickly transmit the more accurate position to the networks.

Many sources whose positions are known precisely will be within the field of view at all times. They will act as fiduciary points and calibrators. Therefore the location accuracy would be largely free of uncorrectable systematic errors.

When not Earth occulted and not too close to the Sun a transient X-ray source or GRB X-ray afterglow will be within the field of view most of the time from birth until it is no longer detectable. The complete history of multiple X-ray afterglows or transients will be observed simultaneously. There is no danger of pile up in the line image or a strong event "blinding" the detector such as the blast that stunned the *Swift* **XRT** on June 21, 2010[30]. The long 1D line image of an intense source is not susceptible to pile-up in the detectors under consideration..

## 4. EXPECTED PERFORMANCE

### 4.1. Telescope Parameters

By ray tracing the performance was simulated of a 1D lobster-eye telescope whose reflectors are deployed along the azimuth of a cylinder with a radius of 1 m with an array of detectors along a concentric cylinder with a radius of 50 cm. This system including the KB module with a focal length of 1.9 m could fit within a 2 m diameter spacecraft such as Orbital Sciences' Taurus, with considerable volume remaining for other instruments, such as an Omni directional gamma-ray detector and the spacecraft systems. However, we did not perform a thorough analysis of structural and thermal factors that might affect these parameters. The mass, volume and power requirements of a cooling system for the solid state detectors were not considered. These factors could require a reduction in collecting area or field of view for compatibility with a 2 m diameter spacecraft.

The lobster-eye mirrors are 9 cm long in one dimension and as high as allowed by the strength of the glass up to 25 cm in the other dimension. The mass estimate is based upon 210 micron thick D263, the same material used by NuSTAR. The lobster-eye mirrors are coated on both faces with 30 nm Ir or Pt plus a 9 nm C overcoat. They are aligned along radii of a 1 m radius cylinder and equally spaced by 6 arc minutes over 180 degrees of azimuth along a 3.14 m arc of a semi-circle. They are divided into twelve 15 degree sectors. The total height of the telescope along the axial (non-focusing) direction would be 1 m divided into four 25 cm sections or a larger number of smaller sections. The focal length of the optics would be 50 cm. The height of the detector array is 50 cm. The field of view is 4 ster with at least 35% of maximum area between polar angles of 45 and 135 degrees with respect to the axis of the cylinder. There is additional field of view outside these angles with less effective area. The optics would be a two dimensional array of 47 lobster-eye mirror modules plus two more that form a KB telescope made of the same materials and the same coating (on only one side) as the lobster-eye mirrors. The coded mask contains a pseudo random distribution of 200 micron wide slits in 100 microns of an absorbing material such as Ni or Ti.

The number glass flats including those of the KB telescope is about 7500, which is comparable to the number of segments in the two NuSTAR telescopes. The mass of the glass mirrors is about 85 kg. In the absence of a mechanical design a rule of thumb suggests that the structure needed to support the flats would raise the total mass by a factor of 1.5 to a total telescope mass of 130 kg.

### 4.2. Effective area

The point response in the focusing dimension shown in Fig. 3 and the effective area were estimated by ray tracing. The transmission of coded mask, which covers the reflectors and its support structure were assumed to be 45% after the fact to account for the loss of open area. X-ray scattering loss by this low surface roughness glass should be negligible. The point response consists of a peak plus a shoulder from rays that pass straight through or are reflected an even number of times.

Table 1. Dimensions assumed for the simulation of the 1D lobster-eye telescope

| Parameter | Numerical Value |
| --- | --- |
| Radius of curvature of the detector array, "R", Fig. 1. | 50 cm, equal to focal length |
| Circumferential size of the detector array | Up to 1.57 m |
| Radius of curvature of the mirror array | 1 m |
| Circumference of the mirror array, (up to 180 deg.) | Up to 3.1 m |
| Total axial length of the mirror array | 1 m |
| Total axial length of the detector array | 0.5 m |
| Azimuth field of view | Up to 180 deg. |
| Polar angle range with > 0.35 maximum area | 45 deg. to 135 deg. |
| Length of the mirrors along a radius | 9 cm |
| Angle and physical space between adjacent mirrors | 6 arc min and 1.75 mm |
| Thickness of the mirrors | 210 microns |
| Reflective coating (on both faces of mirror flats) | 30 nm Ir + 9 nm C |
| Detector, dimensions of a CMOS or CCD chip | 40 mm x 40 mm x 300 microns |
| Number of detector chips | 500 |
| Detector chip packing efficiency | 0.8 |
| Transmission of detector light shield | Same as Chandra ACIS-S |
| Transmission of coded mask plus structure | 0.45 |

The reflectors are distributed uniformly along 180 degrees of azimuth. Therefore the effective area is independent of the azimuth angle except for very small regions at both ends. In the polar direction the maximum effective area occurs when the angle of incidence is normal to the axis of the cylinder. The effective area is shown in Fig. 4 and listed in Table 2 for several energies. The effective area falls to about 0.35 of the maximum when the polar angle is 45 or 135 degrees due primarily to effect of projection, plus a small amount of collimation by the slits of the coded mask, which can be minimized[31].

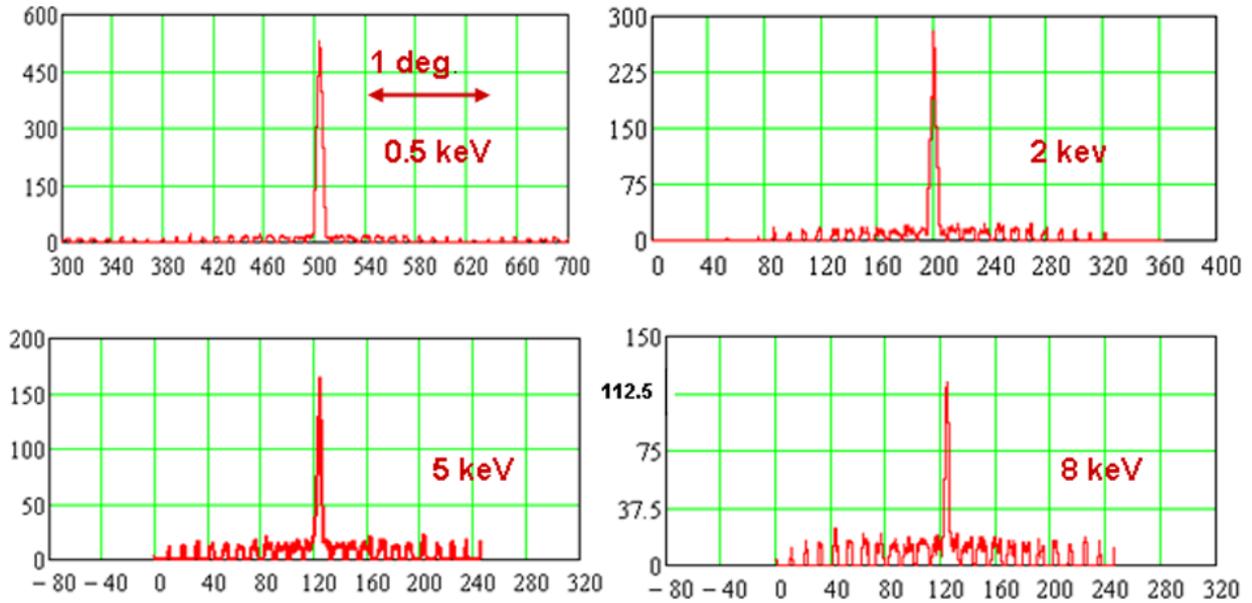

Figure 3. Point response is shown of a 1D lobster-eye telescope for four energies as determined by ray tracing with statistical fluctuations. The horizontal scale is the same for all but their origins differ. The vertical scale is varied so that the peaks will appear at about the same height. The top of the scale is 600, 300, 200, 150 for 0.5, 2, 5, and 8 keV, respectively. The shoulder or halo is due to rays that pass straight through without reflection or are reflected an even number of times.

Assuming their dimensions are 40 mm x 40 mm x 300 microns, the number of CMOS or CCD chips in the array of focal plane detectors is about 500. The KB module requires only a single detector chip, which is set back from the lobster-eye detector array by about 1.4 m more or less according to the spacecraft's ability to accommodate it. By comparison **BAT** of *Swift* has 32,768 CZT chips whose dimensions are 4 mm x 4 mm x 2 mm. The total detector area is about 0.75 m$^2$ in both cases. However, the soft X-ray detectors operate at a low temperature and we have not considered the design of a cooling system.

The angular size of a resolution element is 7 arc min (FWFM) x 90 deg. or 10.5 square degrees. The effect of the shoulders is to increase the solid angle for accepting diffuse X-ray background and the flux of the steady sources by about a factor of 2. The physical size of a resolution element is 0.6 sq. cm... The level of cosmic ray induced background observed in several current missions with X-ray CCD detectors as reported by Hall and Holland[32] indicate that it is small compared to the background from diffuse cosmic X-rays. Therefore, shielding would not be needed.

Fig. 4 is a plot of the effective area as a function of energy for rays normal to the cylinder axis. It includes a factor of 0.45 for the open area of the coded mask and structure as well as a factor of 0.8 for the active area of the array of detector chips.

We assume the sensitive absorption depth of the detectors is 300 microns of Si and that the detector has a light shield with the same X-ray transmission properties as the Chandra ACIS-S's shield, which determines the low energy cutoff. The ray tracing results are shown in Tables 2 & 3. The "Focused Area" in Table 2 is based upon rays that are in the peak of the lobster-eye point response function and are normal to the axis of the cylinder. Those rays are reflected an odd number of times but mostly just once. The coded mask is not present in front of the KB module. An efficiency factor of 0.87 has been applied to account for the opacity of the front and rear aperture support structures that the KB module is likely to need.

As shown in Fig. 3 the point response of the lobster-eye telescope is quite unlike that of the Wolter 1 or KB telescope, where with the aid of internal baffles the number of rays that reach the focal plane without being reflected are eliminated or minimized. In the lobster-eye geometry they are intrinsic in the point response. A ray that is not reflected or reflected an even number of times continues along or nearly along its original path. They account for the shoulder, that is, the broad distribution of rays underlying the peak. Both faces of the mirror reflect which increases the probability that an X-ray will experience multiple reflections. Ray reflected an odd number of times appear in the central peak. Table 3 shows how the power is distributed at 2 keV. The surface brightness of the shoulders alternates between zero and a finite value across an angular range of about 3 degrees. At 5 keV the mean surface brightness of the halo is ~5 % of the central peak.

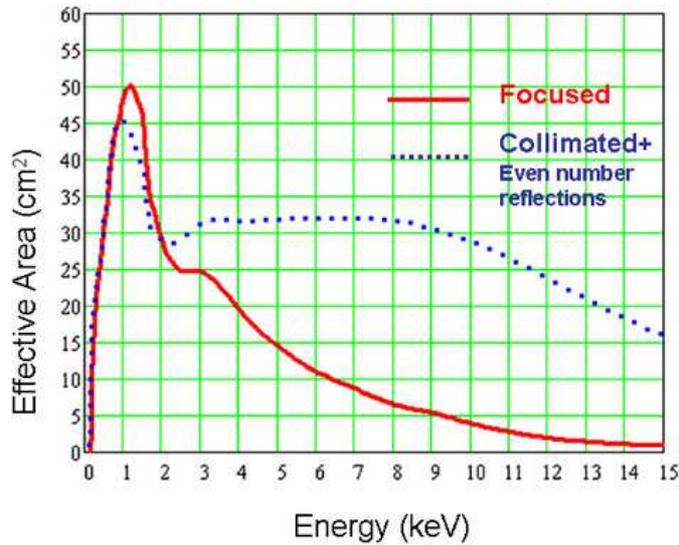

Figure 4. Effective area at a polar angle of 90 degrees is shown as a function of energy for the focused line image and the region defined by the collimating effect of the reflectors, which are X-rays that arrive at detector array without being reflected or are reflected an even number of times.. A detector chip packing efficiency of 80% and the Chandra ACIS-S light shield transmission efficiency have been applied. The transmission of the coded mask plus structure is assumed to be 45%. The blue dotted line is the effective area of the "shoulder" in the point response function shown in Fig. 3.

Table 2. Theoretical Effective Area of Lobster-eye Telescope and KB Module*

| Energy (keV) | Focused Area (cm$^2$) | Collimated Area & 2 reflections (cm$^2$) | Focused + Collimated & 2 refl's. (cm$^2$) | KB Module 2D image (cm$^2$) |
|---|---|---|---|---|
| 0.5 | 30 | 31 | 61 | 110 |
| 0.75 | 42 | 43 | 85 | 166 |
| 1 | 48 | 45 | 93 | 197 |
| 2 | 29 | 29 | 58 | 89 |
| 5 | 15 | 32 | 47 | 30 |
| 8 | 7 | 32 | 39 | 10 |
| 12 | 2 | 24 | 26 | 2 |
| 15 | 1 | 17 | 18 | 0 |

*Includes transmission of a Chandra ACIS-S type light shield, absorption in a 300 micron thick CMOS
 or CCD silicon detector chip with 80% packing efficiency. The transmission of the 1D coded mask is 0.45.

Table 3 Point response of lobster-eye telescope at 2 keV

| | |
|---|---|
| Full angular range of events | 3 degrees |
| Full angular width of central peak | 7 arc minutes |
| Fraction of total power in central peak, below 5 keV | 0.5 |
| FWHM of central peak | 3.5 arc minutes |
| Fraction of total power within FWHM | 0.25 |

According to the ray tracing results the 1D image is a peak with a full width of 7 arc minutes and a FWHM of 3.5 arc minutes plus a shoulder of about 3 degrees.. The shoulders increase the acceptance of diffuse X-ray background and the flux of steady sources that are within the field of view by a factor of 2 up to 5 keV and a larger fraction at higher energies. The precision of the position measurements is the statistical error in determining the centroid of the 3.5 arc minute FWHM central peak. The azimuth angle of sources or bursts with high statistical significance will be measured with sub arc minute precision. Resolving the transient by focusing its rays to a small interval along the azimuth makes determining its position in the polar (axial) direction with the 1D coded mask a relatively simple and rapid process. With a pseudo random distribution of slits that are open along the full range of azimuth the coded mask's properties are independent of the source's azimuth angle

The sub arc minute position found by the 1D lobster-eye/coded mask telescope will be refined to about an arc second by pointing the KB module at it. The KB module will have much finer angular resolution and its area is larger than the lobster-eye's. Therefore its statistical precision in determining the source's position will be much higher.

**4.3 Angular Resolution at Higher Energy with the Coded Mask and "Imaging" Collimators**

The behavior of the coded mask is independent of energy. With a width of 200 microns the angular size subtended by the pseudo randomly distributed slits at a distance of 50 cm from the detector is about an arc minute. It should be possible to determine the polar angular position to better than that using the methods developed for processing coded mask data by *Swift* and *INTEGRAL*. Above 10 keV where the reflection efficiency is low the glass flats are an array of collimators with overlapping fields of view of 0.5 degrees FWHM that are spaced by 0.1 degree around the azimuth of the cylinder. They act as imaging collimators for point sources. The peak intensity occurs at the position of the pair most closely aligned with the source direction. Results from ray tracing 15 keV X-rays from two sources separated by 6 arc minutes are shown in Fig. 5. The sources are clearly resolved and their positions can be determined with arc minute precision. .

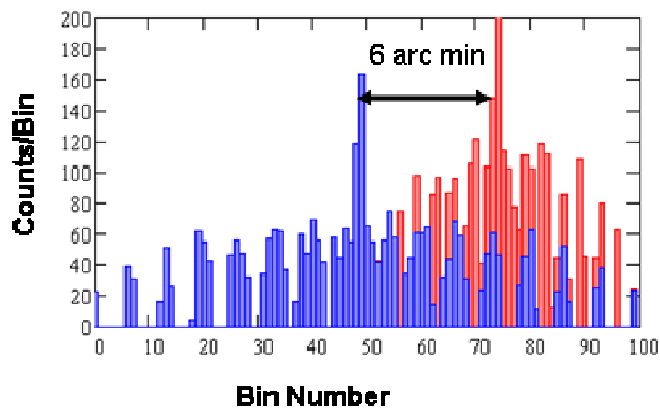

Figure 5. Ray trace images at 15 keV of two point sources of equal intensity separated by 6 arc minutes. Each peak occurs in the direction where the overlapping collimators are most open. There is also a very small contribution from focusing.

# 5. SENSITIVITY

## 5.1. Estimating the sensitivity

The full width of the central peak of the lobster-eye image will be distributed over a region of 7 arc minutes x 90 degrees, or $3.2 \times 10^{-2}$ ster. With the pointing direction fixed we would search for transients by looking for significant increases above the quiescent 0.5 to 8 and 8 to 15 keV rates along the full length of the detector in the polar (axial) dimension on several time scales simultaneously in overlapping successive 8 arc minute bins that step in 4 arc min increments around the azimuth (Sect. 3.2), and. If a significant increase in count rate is detected in one of the 8 arc minute bins the source's azimuth angle will be obtained by calculating the centroid of the focused image and the polar angle will determined by procedures used successfully by *Swift* and ***INTEGRAL*** for obtaining coded mask positions. The effective area for estimating the background is the sum of the "Focused Area" plus "Collimated Area & 2 reflections" that appear in Table 2. The effect of the shoulders in the point response function is to increase the level of diffuse background by a factor of 2. Below 5 keV where its surface brightness is only 4% of the 7 arc minute peak the broad shoulder will not add significantly to the detection sensitivity. However, in the 8 to 15 keV band where the reflection is very low the peak in the shoulder from the imaging collimator effect described in Sect. 4.3 can provide the position and the entire shoulder, useful information on the spectrum and time history.

The background will be larger when the azimuth range of the transient contains another source. With the exception of a few extended sources such as the Cygnus Loop, several supernova remnants and several clusters of galaxies they are mostly point sources, for example, most of the sources in the Uhuru catalog[33]. The presence of a peak of a strong source (intensity > 0.1 Crab Nebula in the Uhuru catalog) will reduce the sensitivity on average by a factor of 18 but will affect less than 1% of the field of view. The shoulders of a strong source will reduce the sensitivity by a factor of 4.4 on average and affect 1/6 at most of the field of view. The rest of the Uhuru sources will reduce the average sensitivity by less than a factor of two in less than 8% of the field of view. The following discussion of the sensitivity applies strictly to the 75 % of the field of view that is not affected on average by the permanent sources.

Table 4 lists the 7 sigma detection sensitivity in terms of $10^{-3}$ of the Crab Nebula (mCrabs) for several observing times. The intensity is assumed to be constant. The putative source has the same spectrum as the Crab Nebula, $10 \times E^{-2}$ photons/cm$^2$-sec and interstellar absorption by a hydrogen column of $3 \times 10^{21}$ H atoms/cm2 along the line of sight. The spectrum of the diffuse cosmic X-ray background is assumed to $10E^{-2.0}$, E < 1 keV, $10E^{-1.4}$ E > 1keV. Although the diffuse cosmic X-ray background is more intense at low energies in some regions of the sky it was not taken into account in the simulation because the assumed Chandra ACIS-S type light shield is opaque below 0.5 keV.

The system has a field of view of 4 ster with at least 35% of the maximum area plus additional solid angle with less area. Not considering possible constraints such as solar avoidance and the need to maintain the detectors at a low temperature but allowing for 10% loss of time to change pointing positions if all points in the sky had equal exposure that time would be 24,760 seconds per 24 hours.

Table 4. Detection sensitivity of constant intensity source relative to $10^{-3}$ Crab Nebula, 0.5 – 8 keV

| **Exposure Time (seconds)** | **7 sigma sensitivity, mCrabs** |
|---|---|
| 30 | 11.5 |
| 100 | 6.2 |
| $10^3$ | 1.5 |
| $10^4$ | 0.62 |
| $2.47 \times 10^4$ (One day in LEO)* | 0.40 |

*Equal exposure for all points in sky and 90 degree polar angle of incidence

According to a sensitivity estimate that takes the 400 second decline time constant into account the supernova found fortuitously by the *Swift* **XRT**[3] would be detected at the 7 sigma level in less than 30 seconds by the hybrid telescope. The on-board computer would recognize the occurrence of the event and point the larger area KB module at it before it is well into the 400 second decline phase.

## 5.2. Comparison with other All-Sky Monitors

### 5.2.1. Current All-Sky X-ray Monitors

The existing all-sky X-ray monitors are the *RXTE* **ASM**, *MAXI*, and the *Swift* **BAT**. **BAT** operates in a higher energy range than the traditional 0.2 to 10 keV X-ray band. It does not detect X-ray sources that do not have a harder component. **BAT** did not detect the supernova flare caught by chance in a field of the *Swift* **XRT**, nor has it detected many examples of the type of X-ray Flashes seen by **BeppoSAX** and **HETE 2.** There also may be "orphan" X-ray afterglows possibly from GRBs that are beamed more narrowly than the afterglow, which are not detected by *Swift*.

Both the *RXTE* **ASM** and *MAXI* are scanning instruments with a much smaller field of view than the lobster-eye telescope. Their exposure time at a given position is relatively short. They are not sensitive enough to detect the supernova seen by the **XRT**. A short lived transient is not likely to occur within their field of view and they would not obtain the complete light curve of the few that do.

The sensitivity of both the *RXTE* **ASM**[34] and *MAXI*[35] is in the range 10 to 20 mCrabs per day. The estimated sensitivity of the hybrid lobster-eye telescope described in this paper (Table 4) is a factor of twenty-five to fifty larger. Furthermore, both the **ASM** and *MAXI* have a limited lifetime and are unlikely to still be operating in several years. However, an instrument similar to the *RXTE* **ASM** will be aboard *ASTROSAT*, a project of the Indian Space Research Organization. The *ASTROSAT* launch date is uncertain.

### 5.2.2. Comparison with a 2D coded mask

All three of the possible very wide field soft X-ray instruments, a 2D coded mask, 1D lobster-eye/1D coded mask hybrid, and a 2D lobster-eye would or could use the same detector system, which is a very large array of position sensitive detectors. Preferably they would be solid state devices such as CCDs, CMOS, or silicon drift, etc. with good spatial and energy providing that they can be maintained at low temperature.

The 1D lobster-eye-coded mask hybrid can be viewed as adding optics to a coded aperture system. The pseudo random distribution of holes would be (but are not required to be) replaced by slits that are open along 180 degrees of azimuth and are pseudo randomly distributed along the cylinder axis. The bare coded mask system would have more effective area but much higher background from the diffuse cosmic X-ray background as well as from all of the discrete sources within its field of view. Both are a much more intense in the soft X-ray band than they are in the hard X-ray band of the *Swift* **BAT**. Several of the known sources are X-ray burst sources, which adds to the difficultly of determining what truly is a supernova or GRB. Comparative estimates of the sensitivity of the 2D coded mask and the lobster-eye/1D coded mask with the same detector system show that the focusing system is several times more sensitive because of the much higher background in the 2D coded mask. Also the coded mask system is not likely to be able to find positions by itself that are accurate to an arc second, the accuracy needed to optically identify signals from GRBs. The 2D coded masks needs to be accompanied by another higher angular resolution X-ray telescope, like the *Swift* **XRT** to refine its positions.

**JANUS** is a very wide field 2D coded mask instrument for the soft X-ray band being studied for a possible Small Explorer mission[36]. The fields of view of **JANUS** and the instrument described in this paper are about the same, 4 ster, and both have a modular configuration. The hybrid system would be more massive but its mass is still within the range of NASA Explorer missions as described in the "Announcement of Opportunity" that was released in November 2010 (Explorer 2011, Solicitation: NNH11ZDA002O).

The effective area of **JANUS** would be much larger but its sensitivity would be much smaller because the background level is much higher. The nominal sensitivity of the hybrid telescope (for sources perpendicular to the cylinder axis) is 11.5 mCrabs, 7 sigma, 30 seconds exposure, (Table 4), while the stated sensitivity of JANUS is 240 mCrabs 7 sigma. However, it is difficult to make a direct comparison because it is not known if the JANUS paper makes the same set of assumptions to estimate sensitivity. **JANUS** devotes some of the spacecraft volume and mass to accommodating an infrared telescope, which presumably could be used to refine positions. However, not every GRB or X-ray Flash will be followed by an infrared afterglow. Ground based, large IR and optical telescopes will be able to perform a much more thorough investigation of the afterglow and its host galaxy than a small telescope aboard a modest spacecraft.

### 5.2.3. Comparison with a 2D lobster-eye telescope

A 2D telescope consists of two orthogonal sets of reflectors that are either integrated as a single optic or are two orthogonal 1D telescopes in series. The geometry of the telescope is spherical. Its effective area, bandwidth, – and background are much smaller than the 1D lobster-eye/coded mask hybrid and its point response function is more complex. At low energies twenty-five per cent of the flux is in each of the following components: a 2D image, two 1D line images, and a halo that is several degrees wide. The 2D lobster-eye has very low background and all sources within the field of view are resolved by focusing. However, the hybrid 1D lobster-eye/coded mask system has a much larger effective area, much larger bandwidth and dynamic range than the 2D lobster-eye. When the 2D telescope is photon limited the hybrid can be very effective. If the event has a very high instantaneous photon flux like the event which occurred on June 21, 2010[29] that overwhelmed the Swift XRT, the 2D lobster-eye could be affected by pile-up in the detector. The consequences are loss of accuracy in measuring positions, fluence, spectra, and timing information. The long 1D line image of an intense source is not susceptible to pile-up.

Several projects are devoted to the development of 2D lobster-eye telescopes for both space and industrial applications. The most interesting approach is a very light weight array of square micro channel plates under study at the Space Research Centre at the University of Leicester[37]. A system with a field of view of 162 x 22.5 degrees (~ 1 ster) and an effective area of about 4 $cm^2$ at 2 keV for the point-like image plus two line images was designed for the **International Space Station (ISS)[38]**. (When scaled up from its 37.5 cm focal length to the 50 cm focal length assumed here that area would become 7 $cm^2$.) The effective area as a function of energy falls very rapidly above 2 keV. The advantages of micro channel plate optics is very low mass and if the figure were perfect, excellent angular resolution.

The dimensions of the Leicester lobster-eye optic are 20 micron square channels with a wall thickness of 4 microns. If the channel plates were slumped perfectly onto a sphere the resolution of the point-like image would be in the range 10 to 20 arc seconds at 1 keV, limited by diffraction and the intrinsic aberrations of the lobster-eye geometry. In practice the FWHM of the Leicester Lobster-ISS telescope point response function was 5 arc min, limited by the difficulty of forming the figure accurately. Coating the micro channel plate walls with a heavy metal for high reflectivity may be an issue. The fraction of the flux that is within the FWHM was not stated. With walls that are only a few microns thick the channel plates will begin to become transparent to X-rays at about 5 keV, which would result in a large increase in the background from discrete sources from many directions.

Other types of 2D lobster-eye telescopes have been described but they do not offer the very low mass advantage of the square micro channel plates[39].

The 1D lobster-eye/coded mask hybrid would have a factor 5 to 10 larger effective area than the 2D lobster-eye telescope, and a much broader bandwidth. It also has a much larger dynamic range. It can detect and find position for bursts or transients with a low photon flux where the 2D telescope would be photon limited and also for very high photon flux events where the 2D's detector is subject to pileup. With larger effective area and broader bandwidth the 1D optic is more capable of measuring the X-ray spectrum of a burst and its X-ray afterglow including the absorption that takes place within the host galaxy. The intrinsic absorption is an indicator of the metal abundances of the distant host galaxy, which is a clue to its star formation history.

A very significant advantage of the 1d lobster-eye hybrid is that with relatively small modifications two modules can be configured as an imaging 2D KB telescope with better angular resolution and more effective area. It can measure positions with arc second precision and low energy spectra more effectively. As more distant GRBs are detected the need to obtain very accurate positions becomes increasingly important. With the 5 arc minute resolution currently being obtained the 2D lobster-eye would not be able to obtain arc second positions on its own.

## 6. SUMMARY AND CONCLUSIONS

The study of soft X-ray variability on all time scales is an important aspect of astronomy. The objectives and capabilities of a soft X-ray telescope system with a very wide field of view would complement those of synoptic telescopes currently being developed for the optical and radio bands. In fact, the soft X-ray band is the most dynamic. It hosts many types of variable galactic and extra-galactic sources. The range of phenomena include: stellar flares particularly from star formation regions, changes in the intensity and spectrum of neutron star and black hole binaries whose behavior tests the laws of general relativity, activity occurring in the vicinity of supermassive black holes, supernova explosions and the X-ray components of GRBs, XRFs, XRRs and their afterglows. Obtaining a precise position from the X-ray afterglow of a GRB led to finding one of the most distant objects in the universe at Z = 8.2. As the host galaxies become more distant the spectrum of a GRB will be increasingly softened by the redshift. At larger distances the redshift increases rapidly with distance resulting in the X-ray component of a GRB becoming an increasingly larger factor in the observed energy spectrum. The search for ever more distant GRBS, etc will benefit considerably by an all-sky X-ray camera that can detect and determine positions for transient soft X-ray events with arc second accuracy. These X-rays measurements and the optical/IR measurements of the afterglow with large telescopes would provide a means of measuring star formation rates and metal abundances as a function of time beginning with emergence from the "dark ages" during the period of reionization.
.

In the soft X-ray band a wide field hybrid focusing 1D lobster-eye telescope-coded mask hybrid with its fine positioning KB telescope is a more powerful tool for detecting, measuring the spectra and finding positions of randomly occurring bursts or transients than either a scanning detector with a limited field of view, a 2D coded mask, or a 2D focusing lobster-eye telescope. Despite its large size the cost of constructing this 1D lobster-eye telescope is relatively low compared to conventional X-ray optics. It is simply a modular array of uniformly spaced identical glass flats that are placed along radii of a cylinder. Very low surface roughness flat glass with a wide range of size and thickness is available commercially at relatively low cost. .A fine positioning 2D KB telescope can be made from the same glass mirrors. Its single chip detector is identical to those of the lobster-eye detector array. All mirrors can be coated with the most highly reflecting materials such as 20 nm iridium with a 9 nm carbon overcoat or even with a more complex multilayer.

The hybrid telescope would offer peripheral vision to, and find new, important targets that the pointed X-ray telescopes with very large effective area and very high resolution spectrometers can study in-depth. The mass and volume requirements of the hybrid system described in this paper are compatible with the type of mission described in the NASA Explorer Announcement of Opportunity issued in November, 2010.